# Fermi-surface of underdoped $LaFeAsO_{1-x}F_x$ as determined by the Haas-van Alphen-effect

G. Li[1], B. S. Conner[1], S. Weyeneth[2], N. D. Zhigadlo[3], S. Katrych[3], Z. Bukowski[3], J. Karpinski[3], D. J. Singh[4], M. D. Johannes[5] & L. Balicas[1]

*’[1]National High Magnetic Field Laboratory, Florida State University, Tallahassee, Florida 32310, USA*

*[2]Physik-Institut der Universität Zürich, Winterthurerstrasse 190, CH-8057 Zürich, Switzerland*

*[3]Laboratory for Solid State Physics, ETH Zürich, CH-8093 Zürich, Switzerland*

*[4]Materials Science and Technology Division, Oak Ridge National Laboratory, Oak Ridge, Tennessee 37831-6114, USA*

*[5] Code 6393, Naval Research Laboratory, Washington, DC 20375, USA*

**Within the Fermi liquid theory for conventional metals, charge-carriers are described as travelling waves (i.e. Bloch states) whose ensemble of allowed wave-vectors describes a three dimensional surface in reciprocal space called the Fermi-surface. From the geometry of this Fermi surface one can infer fundamental properties of any given metal such as the tendency towards an itinerant magnetic instability, associated strength of spin-fluctuations, and the compatible symmetries for a superconducting gap. Here, we present a de Haas-van Alphen (dHvA) effect[1] study on the newly discovered $LaFeAsO_{1-x}F_x$ compounds[2,3] in order to unveil the topography of the Fermi surface associated with their antiferromagnetic and superconducting phases, which is essential for understanding their magnetism, pairing symmetry and superconducting mechanism. Calculations[4] and surface-sensitive measurements[5,6,7] provided early guidance, but lead to contradictory**

**results, generating a need for a direct experimental probe of their bulk Fermi surface. In antiferromagnetic $LaFeAsO_{1-x}F_x$ [8,9] we observe a complex pattern in the Fourier spectrum of the oscillatory component superimposed onto the magnetic torque signal revealing a reconstructed Fermi surface, whose geometry is not fully described by band structure calculations. Surprisingly, several of the same frequencies, or Fermi surface cross-sectional areas, are also observed in superconducting $LaFeAsO_{1-x}F_x$ (with a superconducting transition temperature $T_c$ ~ 15 K). Although one could attribute this to inhomogeneous F doping, the corresponding effective masses are largely enhanced with respect to those of the antiferromagnetic compound. Instead, this implies the microscopic coexistence of superconductivity and antiferromagnetism on the same Fermi surface in the underdoped region of the phase diagram of the $LaFeAsO_{1-x}F_x$ series. Thus, the dHvA-effect reveals a more complex Fermi surface topography than that predicted by band structure calculations[4] upon which the currently proposed superconducting pairing scenarios[10,11,12,13] are based, which could be at the origin of their higher $T_c$s when compared to their phosphide analogs.**

The new iron pnictide superconductors, displaying superconducting transition temperatures as high as 56 K, have some similarities with the cuprate high temperature superconductors, most notably the fact that superconductivity emerges by doping an antiferromagnetic parent compound[14]. On the other hand, the multiband nature of Fe leads to a metallic antiferromagnetic state as opposed to the antiferromagnetic Mott phase seen in undoped cuprates, suggesting a less prominent role for the electronic correlations. Neutron scattering experiments reveal that the antiferromagnetic phase is composed of collinear chains of antiferromagnetically coupled Fe moments[9], presumably of itinerant character[10, 12].

Local Density Approximation (LDA) calculations indicate that the Fermi surface of the non-magnetic compounds is composed of two cylindrical sheets of electron character at the M point of the first Brillouin zone (FBZ) and, depending on the doping level, of at least two more cylinders of hole character at its Γ point[4]; see, for example, Figs. 1 (a) and (b). Such a LDA predicted Fermi surface was indeed observed by both angle resolved photoemission[15] (ARPES) and the de Haas van Alphen effect[16] in the superconducting LaFePO compound. In the undoped (parent) iron arsenides the situation is more complex, since the antiferromagnetic transition is concomitant with a tetragonal to orthorhombic transition[17,18], with certain models suggesting an important role for the orbital degrees of freedom[19, 20]. LSDA Fermi surfaces in these systems are formed in a new Brillouin zone determined by the magnetically ordered and structurally distorted unit cell, see Figs. 1 (c) and (d). Doping carriers into the parent compound suppresses both the antiferromagnetism and the associated structural transition but presumably leaves enough antiferromagnetic fluctuations to trigger superconductivity[10,11,12] as seen at higher doping levels[14,17]. Complex electronic superstructures have been claimed to be relevant for the iron arsenides[21,22,23] with some studies claiming the existence of a pseudogap[24]. However, it is unclear how these structures would alter the geometry of the Fermi surface in the arsenide compounds when compared with their phosphide analogs. ARPES provides contradictory pictures for the Fermi surface of the arsenide compounds, from a general agreement with band-structure calculations[5,6], to evidence for coexisting antiferromagnetic and superconducting orderings[7], to the observation of a new type of “hidden-electronic-order” state deemed the $(\pi,\pi)$ order[25]. These differences might perhaps be attributed to the covalent nature of the La-O bonding since after cleavage the *Ln*OFeAs compounds (where Ln is a lanthanide) expose a polar surface. The concomitant charge redistribution has an electronic structure which deviates strongly with respect to that of the bulk[26].

Given the necessity for a precise determination of the bulk electronic structure at the Fermi level for understanding the superconducting properties of the Fe arsenide compounds, we performed a detailed torque magnetometry study in single crystals of $LaFeAsO_{1-x}F_x$ at very high fields for two doping levels, having respectively antiferromagnetic and superconducting ground states, in order to detect the de Haas van Alphen effect (dHvA). We cannot provide precise values for the fraction $x$ of F doping. But for the first doping level torque magnetometry reveals no trace of superconductivity such as irreversibility or deviations from a paramagnetic response at low temperatures, indicating that it corresponds to an underdoped compound having the aforementioned antiferromagnetic ground state. For the second doping level commercial Squid magnetometry determined a superconducting transition temperature $T_c \approx 15$ K, placing this compound in the underdoped regime but in proximity to antiferromagnetism in their doping-dependent phase-diagram. The magnetic torque, $\boldsymbol{\tau} = \mu_0\, \boldsymbol{m} \times \boldsymbol{H}$, is linearly dependent on $\boldsymbol{m}$ , the bulk magnetization of the sample. Consequently, torque is a bulk thermodynamic physical property in contrast to photoemission. Under a magnetic field, the crossing of the quantized Landau electronic orbits through the Fermi level produces oscillatory components in $\boldsymbol{m}$ , i.e. the dHvA effect, which are periodic in inverse field and whose fundamental frequencies are directly related to the extremal cross-sectional areas of the Fermi surface perpendicular to the field through the Onsager relation[1]. The torque produced by external fields up to 45 T on single-crystals having typical masses of 2 to 3 μg were measured at low temperatures using micro-piezoresistive cantilevers (see, Methods).

Figure 1 presents Fermi surfaces based on density-functional theory (DFT) calculations which were able to capture the essential properties of the Fermi surface of the phosphorus based compounds[16,27] and which are included here as a reference with which to compare our experimental results. It predicts 3 + (2 x 2) = 7 Fermi surface cross-sectional areas for our underdoped and superconducting $LaFeAsO_{1-x}F_x$ compound, with effective masses ranging from 0.46 to 1.7 electronic masses ($m_e$) for the hole orbits and from 1.19 to 1.93 $m_e$ for the electron ones, see Figs. 1 (a) and (b) (see, table SI in Supplementary Information). As for the non-superconducting compound, antiferromagnetism defines a new Brillouin-zone leading to a reconstructed Fermi surface. The precise value of the Fe moment also has a profound influence on the geometry of the Fermi surface. We therefore used a constrained calculation based on a negative U to fix the effective moment at 0.35 $\mu_B$ per Fe as determined by neutron scattering experiments[9]. The resulting Fermi surface contains surfaces with both hole- and electron-character around the Γ point of the Brillouin zone (see, Figs. 1 (c) and (d)), leading to 9 distinct cross-sectional areas (the largest having a dHvA-frequency of 0.713 kT) with masses ranging 0.3 and 2.3 $m_e$ (see, table SII in Supplementary Information).

Figure 2 (a) shows traces of the magnetic torque ***t*** as a function of magnetic field *H*, for a non-superconduting $LaFeAs_{1-x}O_x$ single crystal and for multiple temperatures. One observes a fine reproducible structure superimposed onto a conventional $H^2$ background. Its subtraction exposes this structure, as seen in the lower panel of Fig. 2 (b), where it is plotted as a function of field. This structure is clearly reproducible among the different traces taken at different temperatures. It grows as the field increases and as the temperature is lowered as one would expect for dHvA oscillations. However, the magnitude of this dHvA signal is approximately 3 orders of magnitude smaller than the signal extracted from the LaFePO compound by using a similar set-up[16]. This indicates that the quasiparticle scattering is far more pronounced in the As compounds when compared to their P analogues, although its origin remains unclear. The Fourier transform (FFT) of the dHvA signal as a function of inverse field is shown in Fig. 2 (b). At least 15 peaks are observed in almost all traces acquired at different temperatures, although several of them can be identified as harmonics of more fundamental frequencies. Their large number indicates a severely reconstructed Fermi surface when compared with the 7 frequencies or cross-sectional areas predicted by DFT for the superconducting compound. The observation of frequencies greater than 0.7 kT is at odds with the calculations. The amplitude of several of the FFT peaks is plotted as a function of temperature in Fig. 2 (c). By fitting their temperature dependence to the usual Lifshitz-Kosevich formulae[1], one extracts effective masses ranging from 0.5 to 0.8 $m_e$, again in marked contrast with the DFT calculations (compare, tables SII and SIII in Supplementary Information).

In Fig. 3 (a) we show $\tau$ as a function of the magnetic field for a underdoped and superconducting $LaFeAsO_{1-x}F_x$ ($x$ = 0.1 is its nominal doping level) single crystal and, respectively, for increasing and decreasing field sweeps at a temperature $T$ = 0.5 K and for $\theta = 3^0$. Both traces bifurcate at a field of ~ 25 T, which corresponds to the superconducting irreversibility-field, $H_{irr}$, i.e. the field separating a pinned vortex lattice from either a vortex liquid phase or the metallic state. At very low temperatures $H_{irr}$ is in close proximity or coincides with the upper critical field. Notice the structure superimposed onto $\tau$ which is observed both above and *below* $H_{irr}$. To expose the oscillatory component, seen in the lower panel Fig 2 (a), the background signal was adjusted to a fourth degree polynomial and subsequently subtracted. Fig. 3 (b) shows the Fast Fourier spectrum obtained from both traces when plotted as a function of $H^{-1}$. As one can see, despite the small amplitude of the dHvA signal, the spectrum is quite reproducible and shows a large number of peaks. Remarkably, nearly half of the detected peaks are also observed in the antiferromagnetic and non-superconducting compound (see, tables SIII and SIV in Supplementary Information). Despite the complexity of the FFT spectrum, one can carefully follow the evolution of some of its peaks as a function of $\theta$, as shown in Fig 3 (c). Their amplitude decreases rapidly as the field is oriented towards an in-plane direction. Nevertheless, within the limited angular range where these peaks are observed, they follow the typical dependence expected for two-dimensional Fermi surface cross-sections.

Finally, Fig. 4 displays the FFT spectrum of the dHvA signal from a $LaFeAsO_{0.9}F_{0.1}$ single crystal obtained at different temperatures and at an angle of $13^0$, with the lower panel displaying the temperature dependence of the magnitude of a few peaks. Red lines are fits to the Lifshitz-Kosevich formulae, which leads to effective masses in the order of $2m_e$ for *all* frequencies detected on the upper panel (see, table SIV in Supplementary Information section). All orbits in $LaFeAsO_{0.9}F_{0\,1}$ and in LaFePO have very similar effective masses[16], but the Fermi surface of $LaFeAsO_{0.9}F_{0\,1}$ is clearly composed of a larger number of sheets, implying a higher density of states at the Fermi level. Hence, within the standard BCS theory one could expect a higher $T_c$ for $LaFeAsO_{0.9}F_{0\,1}$ as is effectively seen.

A few of the frequencies, or cross-sectional areas, observed in non-superconducting $LaFeAsO_{1-x}F_x$ correspond to harmonics while others may correspond to magnetic breakdown orbits[1]. Their large number, as well as the pronounced discrepancy between the measured and the calculated effective masses, indicate that the magnetism and concomitant structural distortion has a pronounced effect on the electronic structure at the Fermi level which is not captured by the current weakly-correlated approaches[10,11,12,13].

On the other hand, the observation of a series of frequencies, which are common to both antiferromagnetic metallic and superconducting compounds, would at first glance seem to suggest an inhomogeneous doping distribution where parts of the sample would exhibit the Fermi surface redefined by the antiferromagnetic order. However, in the superconducting compound the increase in their effective masses by a factor of 3 to 4 with respect to the non-superconducting one is inconsistent with this scenario. Instead, the fact that all effective masses in $LaFeAsO_{0.9}F_{0.1}$ have basically the same value, suggests that the same carrier explores pieces of the reconstructed Fermi surface as well as those sheets which are responsible for superconductivity. In other words, it indicates microscopic coexistence between antiferromagnetism and superconductivity as suggested by Ref. 28 and in disagreement with the phase-diagram in Ref. 8. However, neutron scattering indicates that antiferromagnetism completely disappears as the doping increases[9]. This should lead to a Fermi surface reconstruction at a critical doping value. The observed increase in effective masses, by a factor of 3 to 4 for the orbits common to both compounds, when going from the parent to the superconducting compound suggests that such critical doping could correspond to a quantum critical point[29].

**Methods**

Single crystals of $LaFeAsO_{1-x}F_x$ with typical sizes of 60 x 80 x 5 $\mu m^3$, and whose growth is described in detail in Ref. 30, were mounted on micropiezoresistive cantilevers (Seiko PRC 120 or PRC400). Torque magnetometry did not reveal any trace of superconductivity in single crystals in provenance from the batch AN682 used for the present study. Bulk superconductivity was observed in crystals form batch AN690 through torque as well as commercial SQUID magnetometry below $T_c$ ~ 15 K. Single crystals from this batch were selected by measuring the sharpness of their superconducting transition. Torque measurements were performed in a Wheatstone resistance bridge configuration. Devices were placed in a single-axis rotator which was inserted into a $^3$He refrigerator. The whole assembly was inserted into a $^4$He cryostat placed in the bore of either 31 T or 35 T resistive magnets or the 45 T hybrid-magnet at the National High Magnetic field Lab. The magnetic field was swept at a typical rate of 1 T/min. Temperature was stabilized by regulating the pressure of both the $^3$He and the $^4$He baths. For temperatures above 4.2 K, temperature was stabilized by regulating the resistance of a Cernox thermometer. The angle was measured with two Hall probes placed perpendicularly to each other.

<received> Style tag for received and accepted dates (omit if these are unknown).

**Supplementary Information** accompanies the paper on **www.nature.com/nature**.

**Acknowledgements**, The authors acknowledge useful discussion with A. Chubukov, E. Bascones, E. Manousakis and L. P. Gor'kov. LB acknowledges financial support from DOE, Basic Energy Sciences, contract N$^{\underline{0}}$ DE-SC0002613. Work at ORNL was supported by the Department of Energy, BES, Materials Sciences and Engineering. Work at ETH is supported by the Swiss National Science Foundation NCCR MaNEP and project N$^{\underline{0}}$ 124612. The NHMFL is supported by NSF through NSF-DMR-0084173, the State of Florida and DOE.

**Competing Interests statement** The authors declare that they have no competing financial interests.

**Authors' Contributions**. GL, BSC and LB prepared the experimental set-up, performed the measurements and analyzed the data. SW pre-selected some crystals through Squid magnetometry. NDZ, SK, ZB, and JK synthesized and characterized the single crystals by x-ray diffraction measurements. DJS and MJ performed the band structure calculations. LB conceived the project and wrote the manuscript including the input of all co-authors.

**Correspondence** and requests for materials should be addressed to L.B. (balicas@magnet.fsu.edu).

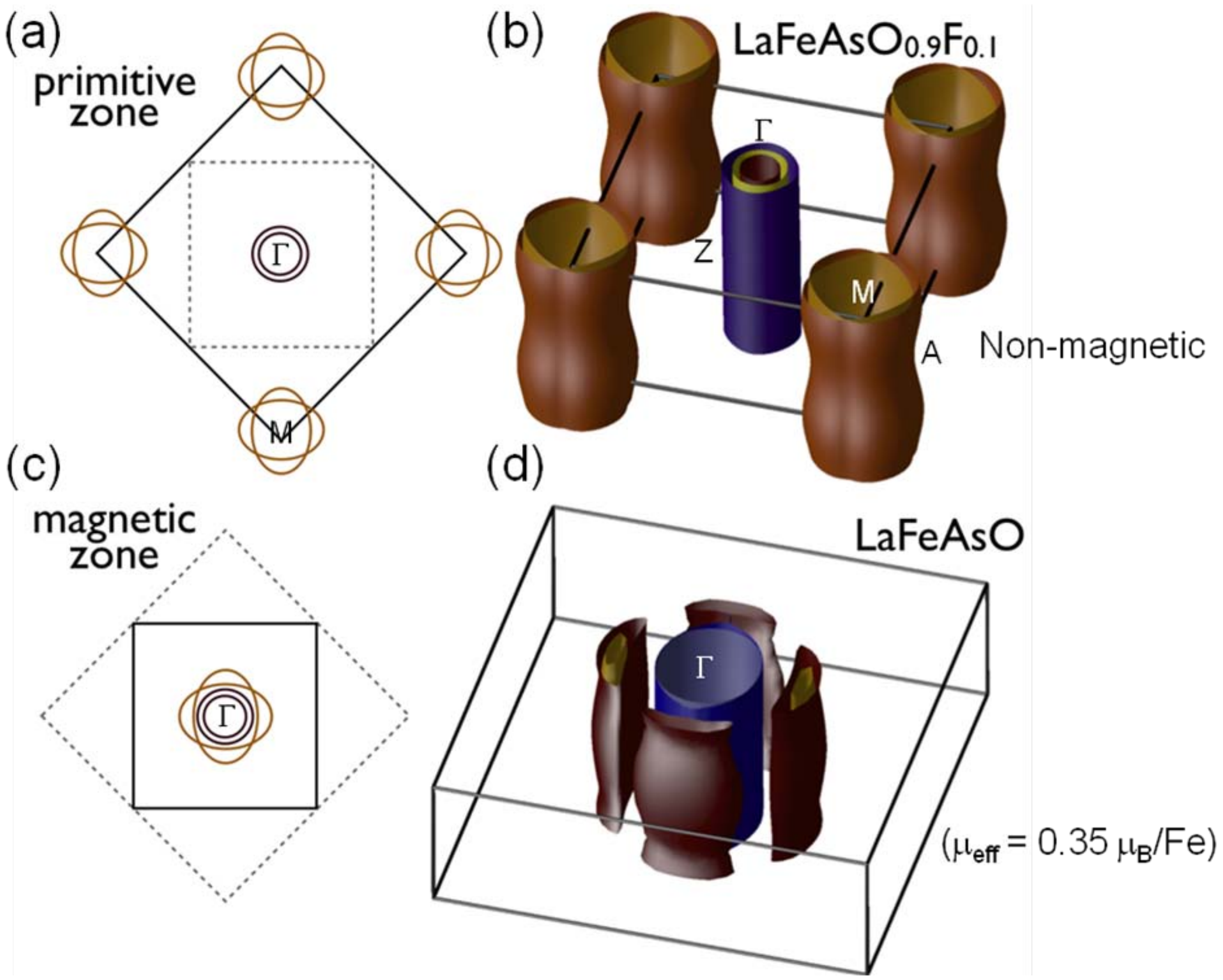

<FIGURE 1> (a) Top view of the Fermi surface cross-sectional areas obtained by Local Density Approximation calculations (LDA) for superconducting $LAFeASO_{0.9}F_{0.1}$ (see Supplementary Information section). The starting point for these calculations is the crystallographic structure of single-crystals in provenance of the same batches used for our torque magnetometry study, as determined by single crystal x-ray diffraction measurements. The Fermi surface is composed of two hole-like cylinders around the Γ-point of the Brillouin zone and of two larger electron-like cylinders around the M-point. Solid line depicts the respective two-dimensional Brillouin zone while dotted line defines the magnetic Brillouin zone resulting from the collinear magnetic order observed below $T_N \cong 137$ K [9]. (b) A three-dimensional perspective of the same Fermi surface sheets indicating their respective degrees of warping along the $k_z$ direction. (c) Top view schematic of the Fermi surface cross-sectional areas in (a), folded into Brillouin zone. Both hole- and electron-like Fermi surfaces are now located around the Γ point. (d) The calculated Fermi surfaces of LaFeAsO as obtained by LSDA+U calculations, adjusting U to yield a value $\mu_{eff} = 0.35\ \mu_B$ for the effective magnetic moment of Fe, as measured by neutron scattering experiments[9].

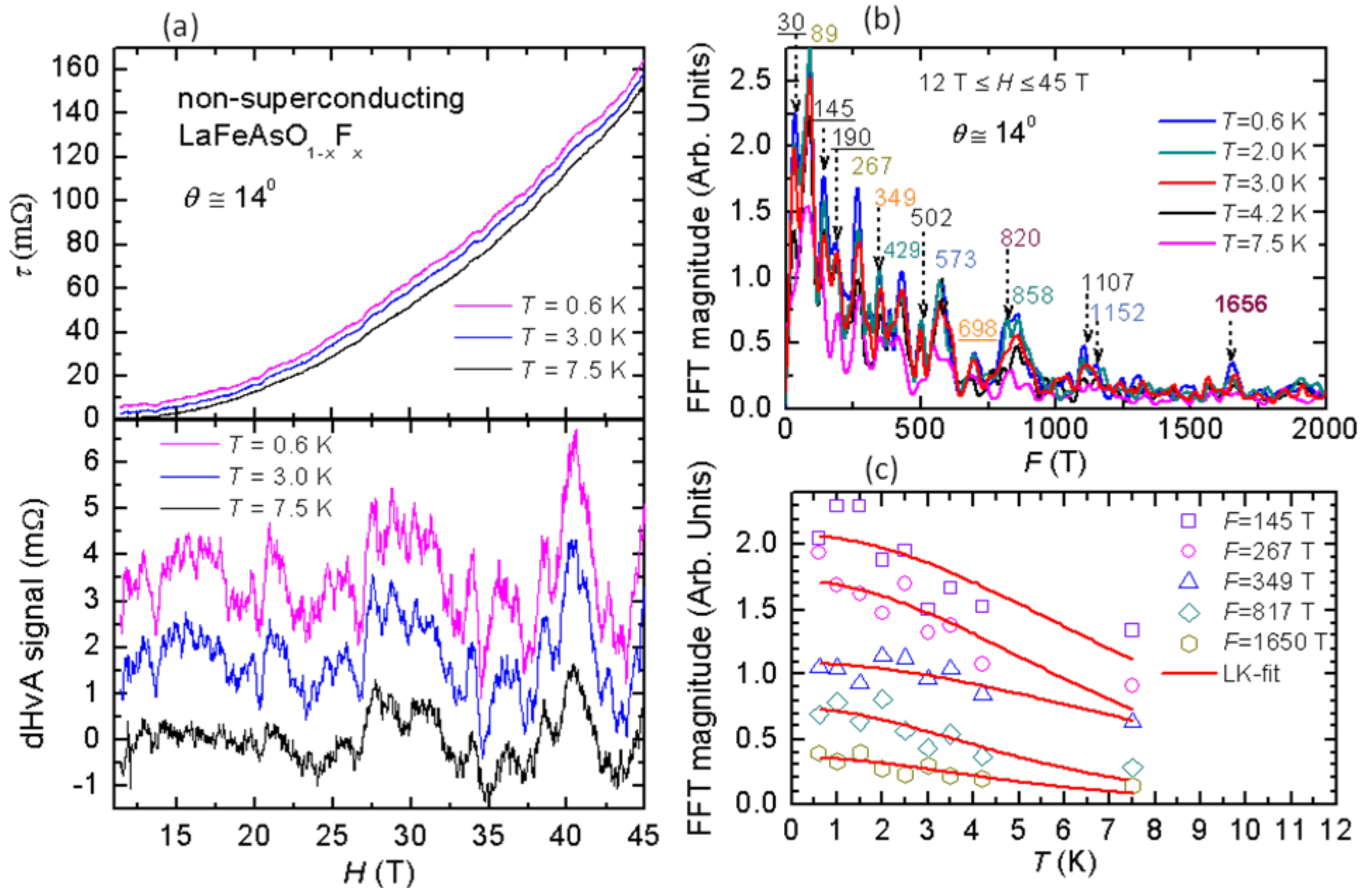

<FIGURE 2> (a) Top panel: the magnetic torque $\tau$, as a function of magnetic field $H$, for several temperatures for an antiferromagnetic and non-superconducting $LaFeAsO_{1-x}F_x$ single crystal. Here, the angle $\theta$ between the magnetic field and the inter-planar direction is $14^0$. Curves are vertically displaced for clarity. Bottom panel: The complex oscillatory component, i. e. the de Haas van Alphen effect, superimposed onto the torque signal shown in (a) and obtained by subtracting a $H^2$ background. The different curves are vertically shifted for clarity. (b) The corresponding fast Fourier transform (FFT) of the dHvA signal for several temperatures. The hybrid magnet is composed of an external superconducting solenoid that provides a background field of ~ 11. 5 T, and a resistive solenoid placed inside its bore which provides the extra field up to 45 T. During the measurements the superconducting solenoid is kept at constant maximum field while one sweeps the resistive magnet. Consequently, the field range between 12 to 45 T was used to extract the FFT spectrum. Several peaks are detected and indicated in the figure, but only a small fraction of these match values predicted by the LDA calculations. The colour code indicates peaks which are interrelated as harmonics. (c) Temperature dependence of the magnitude of a few of the observed FFT peaks. Red lines are fits to the Lifshitz-Kosevich expression $y/\sinh y$, with $y = 14.69\, m^* T/H$, from which one extracts the effective masses $m^*$ of the charge carriers. All the extracted effective masses have values between 0.5 and 0.8 of the mass of a bare electron.

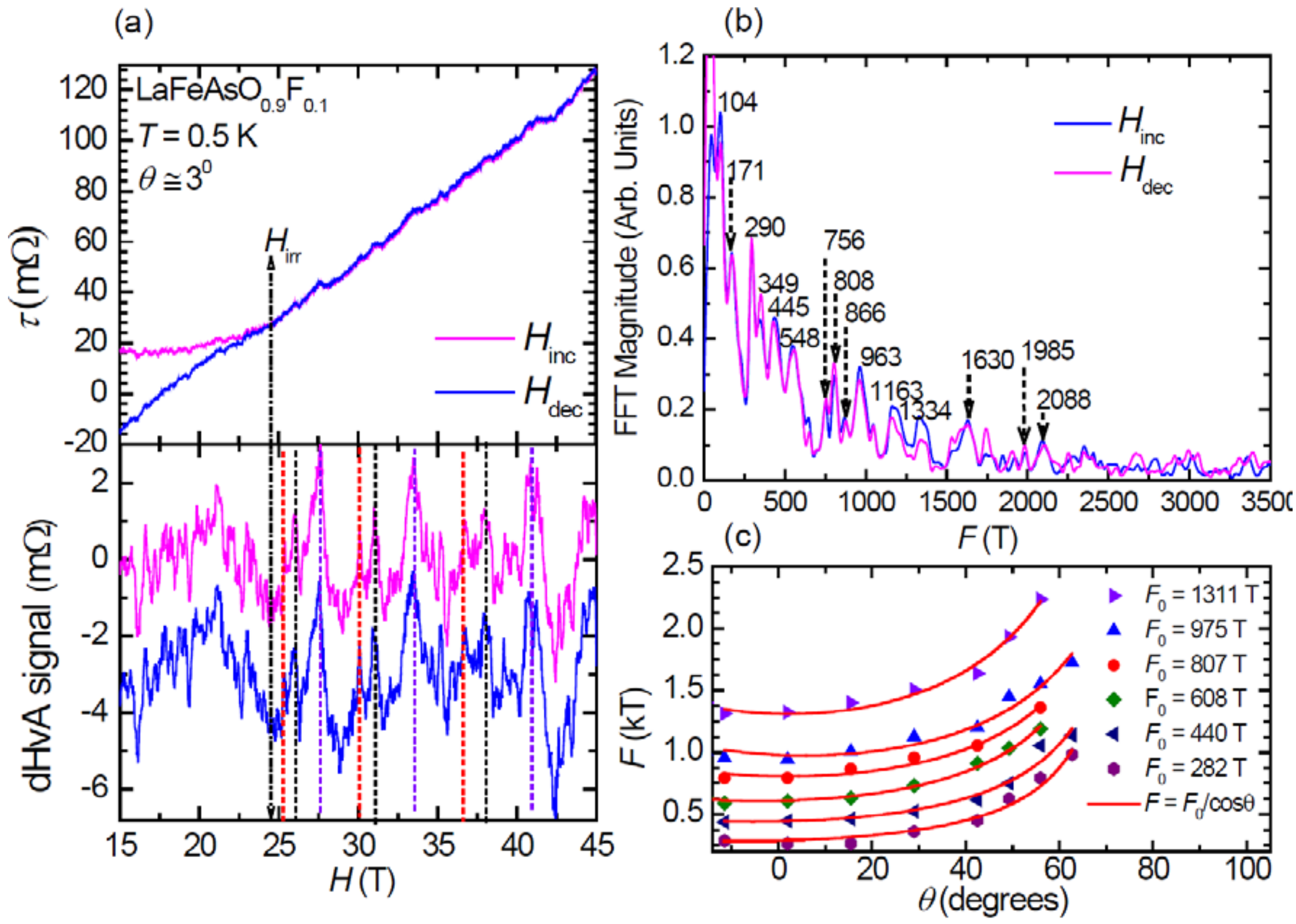

<FIGURE 3> (a) Top panel: Magnetic torque $\tau$ for a single crystal of superconducting $LaFeAsO_{0.9}F_{0.1}$, as a function of both increasing and decreasing magnetic fields and for a temperature $T \cong 0.5$ K. The superconducting irreversible field, $H_{irr} \cong 24.5$ T, is indicated by a vertical dotted line. The vertical coloured dotted lines indicate an oscillatory pattern that repeats itself at increasing field intervals (or periodic in inverse field). Bottom panel: The dHvA signal of both torque traces shown in (a) and obtained by subtracting a quadratic in field background. For the sake of clarity, the decreasing field sweep curve is vertically shifted. (b) The corresponding fast Fourier transform (FFT) of both dHvA signals indicating that a number of peaks are observed in both traces although with slightly different relative amplitudes. The field range from 15 to 45 T was used to extract the FFT spectrum. (c) Angular dependence of a few of the peaks observed in (b). Red lines are fits to the expression $F = F_0/\cos\theta$, which is the angular dependence expected for cylindrical Fermi surface cross-sectional areas.

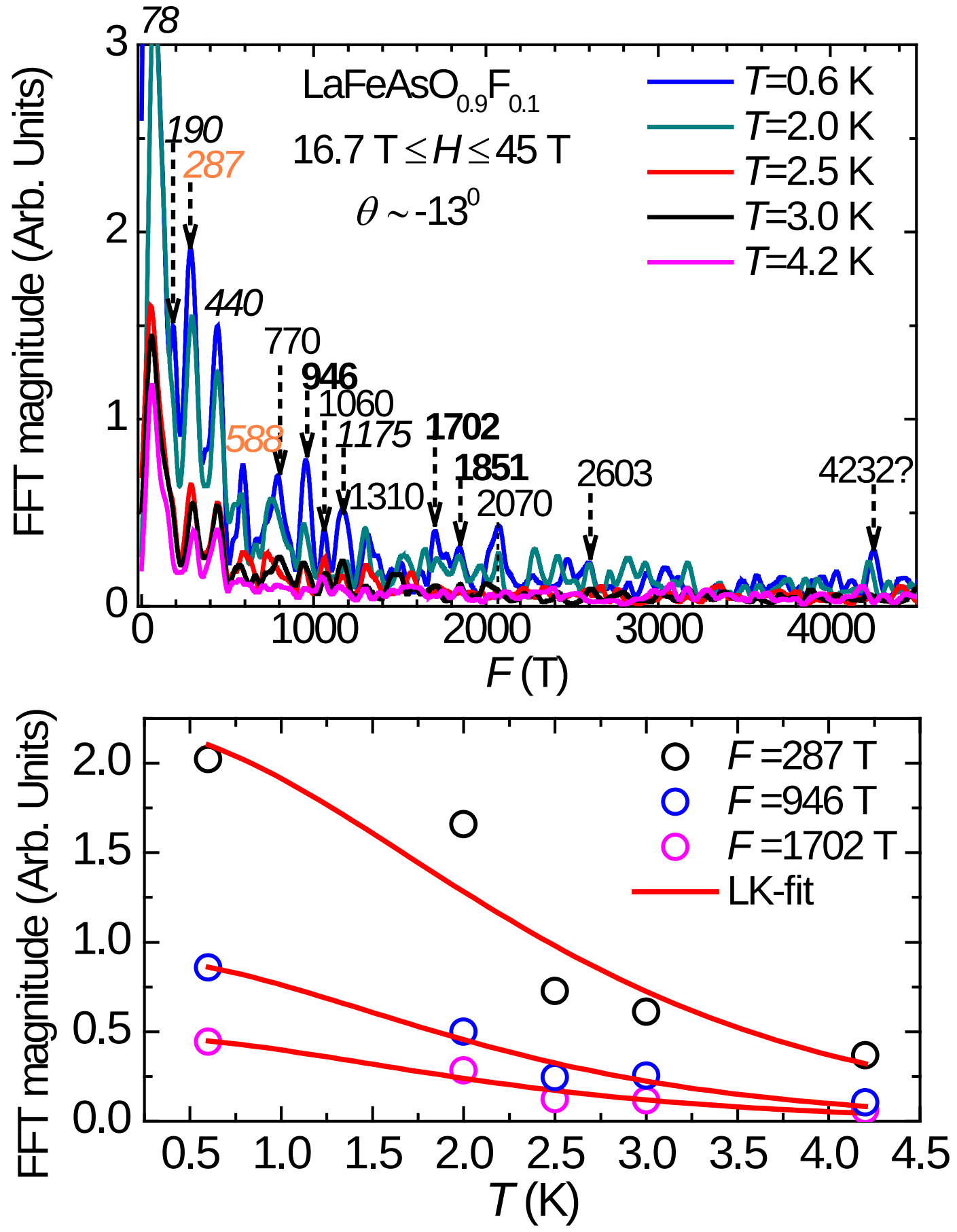

<FIGURE 4> Top Panel: The FFT spectra of the dHvA signal measured at several temperatures for a $LaFeAsO_{0.9}F_{0.1}$ single crystal. To extract the FFTs the field range between 16.7 and 45 T was used. Several peaks are detected and indicated in the figure, but the ones whose frequency values are given by underlined numbers were predicted by the LDA calculations. Italicized numbers indicate peaks/frequencies previously observed in the undoped LaFeASO compound.  Finally, the colour code indicates peaks which are interrelated as harmonics. Bottom panel: Temperature dependence of the magnitude of a few of the observed FFT peaks. Red lines are fits to the Lifshitz-Kosevich expression $y/\sinh y$, with $y = 14.69\, m^* T/H$, from which one extracts effective masses $m^*$ of approximately 2 electron masses for all frequencies.

## Supplementary information

### Band Structure Calculations

All density functional theory Fermi surfaces are calculated with the Wien2k augmented plan wave plus local orbital (APW+lo) code [S1], using the local density approximation (LDA) [S2] and its extension, LDA+U [S3] (discussed below), and with the experimental lattice parameters and atomic positions. Magnetic Fermi surfaces are calculated in the so-called "stripe" phase in which the basic unit cell is doubled both in-plane and along the c-axis to account for the anti-ferromagnetic ordering in both directions; the magnetic space group is *Pccm* (#49). We use a mesh of 23x23x7 kpts in the irreducible Brillouin zone for self-consistency and a mesh of 49x49x16 to produce smooth Fermi surfaces for the purpose of calculating orbits and effective masses. For the doped (superconducting) materials, we apply the virtual crystal approximation (VCA) at the oxygen site. In the VCA, a pseudo-atom with increased core charge and electrons equal to the extent of doping in the material is substituted at the dopant site. In the case studied here, we use a pseudo atom of charge 8.1 at the oxygen site to mimic the (nominal) F doping. For these compounds, we do a non-magnetic calculation in the tetragonal unit cell *P4/nmm* (#129) with a mesh of 23x23x10 kpts for self-consistency and a mesh of 101x101x47 for the Fermi surfaces used to calculate frequencies and effective masses.

Because the Fermi surfaces are highly dependent on the local Fe moment which is strongly exaggerated by LDA ($\mu_{calc} = 1.78\ \mu_B$), we find little agreement between calculated and measured frequencies or masses by simply using LDA. To rectify this, we apply the LDA+U methodology with a negative value of U. The LDA+U scheme applies a Hubbard-like U in a mean field way to a specified set of orbitals in a given system.

**Table SI. Calculated Fermi surface cross sectional areas, corresponding dHvA frequencies, and effective masses for $LaFeAsO_{0.9}F_{0.1}$**

| Cross sectional area | *F* (k T) | $m/m_e$ |
|---|---|---|
| 3Γ | 1.37 | 1.43 |
| 2Γ | 0.68 | 1.7 |
| 1Γ | 0.38 | 0.46 |
| 3Z | 1.37 | 1.43 |
| 2Z | 0.68 | 1.7 |
| 1Z | 0.38 | 0.46 |
| 4M | 2.16 | 1.93 |
| 5M | 2.15 | 1.49 |
| 4A | 2.82 | 1.19 |
| 5A | 2.15 | 1.49 |

**Table SII. Calculated Fermi surface cross sectional areas, corresponding dHvA frequencies, and effective masses for LaFeAsO**

| Cross-sectional areas for a reduced moment of 0.35 $\mu_B$ | *F* (kT) | $m/m_e$ |
|---|---|---|
| 2Γ | 0.699 | 1.06 |
| 1Γ | 0.614 | 0.70 |
| 2Z | 0.713 | 1.13 |
| 1Z | 0.631 | 0.68 |
| 2Xa | 0.222 | 0.94 |
| 1X | 0.02 | 0.3 |
| 2Ma | 0.157 | 0.79 |
| 1M | 0.056 | 0.43 |
| 2Xb | 0.184 | 1.64 |
| 2Mb | 0.059 | 2.3 |

One effect of adding the LDA+U term is that the local magnetic moment is increased. Conversely, applying a negative U has the effect of decreasing the local moment. Though there is no true underlying physical justification for this procedure (the actual missed physics is likely spin fluctuations which are not at all captured by either LDA or LDA+U), the role of the moment in the electronic structure is so strong that depressing it to the observed value by whatever method necessary allows a much closer comparison between theory and experiment [S4,S5,S6,S7]. We use a value of U = - 1.9eV in the fully localized limit [S8] which results in a moment of $\mu = 0.35\ \mu_B$.

**Table SIII. Experimentally determined dHvA frequencies and effective masses for underdoped and non-superconducting $LaFeAsO_{1-x}F_x$, with possible harmonics in blue.**

| Non-SC $LaFeAsO_{1-x}F_x$ – Experimental | |
|---|---|
| $F$ (kT) | $m/m_e$ |
| 0.03 ± 0.01 | 0.9 ± 0.2 |
| 0.09 ± 0.01 | 0.61 ± 0.07 |
| 0.15 ± 0.01 | 0.5 ± 0.1 |
| 0.19 ± 0.01 | 0.5 ± 0.1 |
| 0.27 ± 0.02 | 0.63 ± 0.09 |
| 0.35 ± 0.02 | 0.48 ± 0.09 |
| 0.43 ± 0.02 | 0.56 ± 0.09 |
| 0.50 ± 0.02 | 0.6 ± 0.1 |
| 0.57 ± 0.02 | 0.7 ± 0.1 |
| 0.7 ± 0.02 | 0.74 ± 0.13 |
| 0.82 ± 0.02 | 0.85 ± 0.14 |
| 0.86 ± 0.02 | 0.7 ± 0.1 |
| 1.1 ± 0.02 | 0.67 ± 0.16 |
| 1.15 ± 0.02 | 0.7 ± 0.1 |
| 1.66 ± 0.02 | 0.87 ± 0.15 |

**Table SIV. Experimentally determined dHvA frequencies and effective masses for underdoped and superconducting $LaFeAsO_{1-x}F_x$, with possible harmonics in blue, and in red those frequencies observed also in the antiferromagnetic metallic phase of $LaFeAsO_{1-x}F_x$**

| $LaFeAsO_{0.9}F_{0.1}$ - Experimental | |
|---|---|
| $F$ (kT) | $m/m_e$ |
| 0.08 ± 0.02 | 1.9 ± 0.3 |
| 0.19 ± 0.01 | 2.2 ± 0.2 |
| 0.29 ± 0.01 | 2.0 ± 0.2 |
| 0.44 ± 0.01 | 1.8 ± 0.2 |
| 0.58 ± 0.04 | 1.9 ± 0.3 |
| 0.77± 0.04 | 2.1± 0.3 |
| 0.95 ± 0.01 | 2.3 ± 0.2 |
| 1.06 ± 0.01 | 1.7 ± 0.3 |
| 1.18 ± 0.01 | 2.0 ± 0.3 |
| 1.3 ± 0.02 | 1.6 ± 0.5 |
| 1.7 ± 0.01 | 2.3 ± 0.2 |
| 1.85 ± 0.01 | 1.9 ± 0.4 |
| 2.1 ± 0.07 | 2.3 ± 0.3 |
| 2.6 ± 0.02 | 2.0 ± 0.4 |

***iv*) References**